\documentstyle[12pt]{article}
\def\be{\begin{equation}}
\def\ee{\end{equation}}
\def\ba{\begin{array}}
\def\ea{\end{array}}
\def\bea{\begin{eqnarray}}
\def\eea{\end{eqnarray}}

\begin{document}
\begin{center}
{\Large \bf Structure effects in the region of superheavy elements via the
$\alpha$-decay chain of $^{293}$118} \\

\bigskip

{\bf Raj K. Gupta$^{1,2}$, Sushil Kumar$^{1}$, Rajesh Kumar$^{1}$, \\
M. Balasubramaniam$^{1,2}$,
\\
and \\
W. Scheid$^2$}

\medskip

$^1$ Physics Department, Panjab University, Chandigarh- 160014, India\\
$^2$ Institut f\"ur Theoretische Physik, J.-L.-Universit\"at, Heinrich-Buff-Ring 16, D-35392 Giessen, Germany\\

\medskip

\end{center}

\vspace*{0.3cm}

\begin{center}
{\bf Abstract}
\end{center}

\medskip

\baselineskip 18pt

The $\alpha$-decay chain of $^{293}$118, first proposed in the
Berkeley cold fusion experiment $^{208}$Pb($^{86}$Kr,1n) and now
retracted, is calculated by using the preformed cluster model
(PCM) of one of us (RKG). Also, the
possible branchings of $\alpha$-particles to heavier cluster
decays of all the parents in this chain are calculated for the
first time. The calculated Q-values, penetrabilities and
preformation factors for $\alpha$-decays suggest that the
$^{285}$114 nucleus with Z=114, N=171 is a magic nucleus, either
due to the magicity of Z=114, or of N=172 or of both. The N=172 is
proposed to be a magic number in certain relativistic mean-field
calculations, but with Z=120. The calculated cluster decays point
to new interesting possibilities of $^{14}$C decay of the
$^{281}$112 parent, giving rise to a (reasonably) deformed Z=106,
N=161, $^{267}$106 daughter (N=162 being now established as the
deformed magic shell) or to a doubly magic $^{48}$Ca cluster
emitted from any of the parent nucleus in the $\alpha$-decay
chain. Apparently, these are exciting new directions for future
experiments.\\

\vfill\eject

The synthesis of Z=118 element in the cold fusion reaction $^{208}$Pb($^{86}$Kr,1n)
via the observed $\alpha$-decay chains had created much excitement recently.
This reaction was first made at Berkeley \cite{ninov99}, establishing three
decay chains, with a resulting very high fusion cross section of $2.2^{+2.6}_{-0.8}$ pb
compared to the limiting value of $\sim 1$ pb for the cold fusion reactions
leading to the heaviest Z=112 element. However, this experiment is now
retracted \cite{ninov02,loveland00}, since many other subsequent attempts
\cite{hofmann00,morimoto01,stodel01,grego00} at various other laboratories
(GSI,RIKEN,GANIL) around the World failed to reproduce these data and "one
event" upper limit of $<$0.5 pb has now been put for this reaction. Such a
resulting situation has some consequences for what has been the cause for a
large excitement for nuclear structure studies in the region of superheavy
elements \cite{rutz97,patra97,bender99,patra99,patra2k,gupta01,beckmann02},
discussed below.

The measured large cross section for Z=118 element, now retracted, was considered
\cite{gupta01} as a possible signature of our approaching the centre of
island of stability for superheavy elements (SHE) around Z=120, predicted by
the well founded relativistic mean-field (RMF) calculations
\cite{rutz97,patra97,bender99,patra99,patra2k}. However, the lowering down of
the fusion cross section for this reaction to $<$0.5 pb means that we must go
up the ladder of SHE rather steadily, as was proposed by another calculation
of some of us \cite{gupta97}, based on the well accepted Quantum Mechanical
Fragmentation Theory \cite{gupta99a,maruhn74,gupta75,gupta77a,gupta77b,gupta77c}.
The surprises, if any, were expected to lie in the overshooting of this centre
of island of SHE by means of (neutron-rich) radioactive nuclear beams for the
magic N=184. Then the question would arise for protons, whether Z=110 or 114
is magic, as has been predicted since the early days of this subject, or it
is around Z=120, as is predicted more recently by the above mentioned RMF
calculations.

In this paper, we attempt to look for an answer to the question raised in the
last paragraph above, regarding the nuclear structure effects in SHE, as well
as to the cause for the failure of Z=118 experiment. We do this by analysing
theoretically the $\alpha$-decay chain for $^{293}$118. This is only an
exploratory study and can be extended to other heavy nuclei in this region.
Also, we have calculated for the first time the possible branching of
$\alpha$-decay to any heavy cluster decay, at any stage of the $\alpha$-decay
chain of $^{293}$118. Such a process of, so-called, cluster radioactivity
should open new vistas for the decay studies of SHE, with a possible landing
at some new or known magic daughter nucleus. We have used here for our decay
calculations the Preformed Cluster Model (PCM) of Gupta
\cite{gupta88,malik89,gupta99b}.

The preformed cluster model (PCM) uses the dynamical collective coordinates of
mass and charge asymmetries
$$\eta={{(A_1-A_2)}/{(A_1+A_2)}}$$
and
$$\eta _Z={{(Z_1-Z_2)}/{(Z_1+Z_2)}},$$
first introduced in the Quantum Mechanical Fragmentation Theory
\cite{gupta99a,maruhn74,gupta75}. These are in addition to the usual
coordinates of relative separation R and deformations $\beta_i$ ($i=1,2$).
Then, in the standard approximation of decoupled R- and $\eta $-motions
\cite{gupta88,malik89,gupta99b,gupta91}, the decay half-life $T_{1/2}$ or the
decay constant $\lambda $ in PCM is defined as
\be
\lambda ={{{ln 2}\over {T_{1/2}}}}=P_0\nu _0 P.
\label{eq:1}
\ee
Here $P_0$ is the cluster (and daughter) preformation probability and P the
barrier penetrability which refer, respectively, to the $\eta$ and R motions.
The $\nu _0$ is the barrier assault frequency. The $P_0$ are the solutions of
the stationary Schr\"odinger equation in $\eta$,
\be
\{ -{{\hbar^2}\over {2\sqrt B_{\eta \eta}}}{\partial \over {\partial
\eta}}{1\over {\sqrt B_{\eta \eta}}}{\partial\over {\partial \eta
}}+V_R(\eta )\} \psi ^{({\nu})}(\eta ) = E^{({\nu})} \psi ^{({\nu})}(\eta ),
\label{eq:2}
\ee
which on proper normalization are given as
\be
P_0={\sqrt {B_{\eta \eta}}}\mid \psi ^{({0})}(\eta (A_i))\mid ^2\left ({2/A}\right ),
\label{eq:3}
\ee
with i=1 or 2 and $\nu$=0,1,2,3.... We are interested here only in the ground
state solution ($\nu$=0) since the $\alpha$ (as well as the proposed heavy
cluster) emissions in the considered decay chain are the ground state decays.
Eq. (\ref{eq:2}) is solved at a fixed $R=R_a=C_t(=C_1+C_2)+d$, the first
turning point in the WKB integral for penetrability P (see Fig. 1 and Eq.
\ref{eq:5}), since this value of R (instead of $R=R_0$, the compound nucleus
radius) assimilates to a good extent the effects of both the deformations of
two fragments and neck formation between them \cite{kumar97}. In this way,
the two-centre nuclear shape is simulated through a neck-length parameter $d$
added to scission configuration, which for actinides is nearly zero
\cite{kumar97}, and is taken to be so for superheavy nuclei. The role of
deformation in the scattering potential V(R) is shown \cite{kumar97} to lower
the interaction barriers but not the relative formation yields. The $C_i$ are
S\"ussmann central radii $C_i=R_i-({1/R_i})$, with the radii
$R_i=1.28A_i^{1/3}-0.76+0.8A_i^{-1/3} fm$.

The fragmentation potential $V_R(\eta )$ in (\ref{eq:2}) is calculated simply
as the sum of the Coulomb interaction, the nuclear proximity potential
\cite{blocki77} and the ground state binding energies of two nuclei,
\be
V(R_a, \eta,\eta_Z) = \frac{Z_{1} Z_{2} e^{2}}{R_a}-
\sum_{i=1}^{2} B(A_{i}, Z_{i}, \beta _{i}) + V_{P},
\label{eq:4}
\ee
with B's taken from the 1995 experimental compilation of Audi and Wapstra
\cite{audi95} and from the 1995 calculations of M\"oller et al.
\cite{moeller95} whenever not available in \cite{audi95}. Thus, full shell
effects are contained in our calculations that come from the experimental
and/or calculated \cite{moeller95} binding energies.
The shell effects in the calculated binding energies of M\"oller et al.
\cite{moeller95} are obtained in Strutinsky way \cite{strutin67} by using
the folded-Yukawa single-particle potential and macroscopic finite-range
droplet model (FRDM). The model parameters are fitted to the ground-state
masses of 1654 nuclei, ranging from $^{16}O$ to $^{263}106$, and to 28
fission-barrier heights. Hence, their extrapolation to heavier elements,
studied here in this paper, is expected to be a realistic one. Note that
the familiar magic numbers N=Z=50, 82 and N=126 are given by these
calculations and there is {\it no} assumption made about the magic numbers
in the extrapolated region of SHE. The center of the superheavy region in
these calculations is found to be located at $^{294}115_{179}$
\cite{moeller95}.

The charges Z$_1$ and Z$_2$ in (\ref{eq:4}) are fixed
by minimizing the potential in $\eta_Z$ coordinate, which automatically
minimizes the $\beta_i$ coordinates. Note that the minimized $\beta_i$'s are
{\it not} always for the spherical nuclei since the total binding energy
$B_1+B_2$ of the decay products is minimized and not their individual
$B_1$ or $B_2$. The Coulomb and proximity potentials in (\ref{eq:4}) are
for spherical nuclei. The mass parameters $B_{\eta \eta}(\eta )$, representing
the kinetic energy part in (\ref{eq:2}), are the classical hydrodynamical
masses \cite{kroeger80}. The shell effects in masses are shown \cite{gupta99b}
not to affect the order of calculated preformation yields.

The WKB tunnelling probability, calculated for the tunneling path shown in
Fig. 1, is  $P=P_i P_b$ with
\be
P_i=exp[-{2\over \hbar}{{\int }_{R_a}^{R_i}\{ 2\mu [V(R)-V(R_i)]\}
^{1/2} dR}]
\label{eq:5}
\ee
\be
P_b=exp[-{2\over \hbar}{{\int }_{R_i}^{R_b}\{ 2\mu [V(R)-Q]\} ^{1/2} dR}].
\label{eq:6}
\ee
These integrals are solved analytically \cite{malik89} for $R_b$, the second
turning point, defined by $V(R_b)=Q$-value for the ground-state decay.

The assault frequency $\nu _0$ in (\ref{eq:1}) is given simply as
\be
\nu _0=(2E_2/\mu )^{1/2}/R_0,
\label{eq:7}
\ee
with $E_2=(A_1/A) Q$, the kinetic energy of the lighter fragment, for the
$Q$-value shared between the two products as inverse of their masses.
Eq. (\ref{eq:7}), used here, results in  $\nu _0\approx 3\times 10^{21} s^{-1}$,
whereas the more often used value in literature is $\sim 10^{22} s^{-1}$ for
even parents and $\sim 10^{20} s^{-1}$ for odd parents \cite{gupta94}.

Figure 2 shows the calculated (logarithms of) $\alpha$-decay half-lives,
$log_{10}T^{\alpha}_{1/2}$ (s), as a function of the masses of parent nuclei
for the whole decay chain of $^{293}$118. Also shown in Fig. 2 are the results
of another recent calculation by Royer \cite{royer00} based on the generalized
liquid drop model (GLDM) with the binding energies for Q-values taken from the
Thomas-Fermi (TF) model \cite{myers96}. The TF model uses the well known
Seyler-Blanchard effective nucleon-nucleon interaction where the
momentum-dependent and density-dependent terms are also included. The model
parameters are fitted again to ground-state masses of 1654 nuclei with
N,Z$\ge$8 and the 40 fission-barrier heights.

We notice in Fig. 2 that the calculated numbers for both the models present
an interesting result: the $\alpha$-decay half-life for Z=114, A=285
is very high, which means that the parent nucleus $^{285}_{171}$114 is very
stable against $\alpha$-decay. This stability can be attributed to either the
magicity of protons at Z=114 or of neutrons at $N\approx 172$ or to both,
perhaps more so to Z=114 since N=172 is predicted to be magic only when Z=120
\cite{rutz97,patra97,bender99,patra99,patra2k,beckmann02}. The predicted
half-lives in the two models differ by about four orders of magnitude, which
is partly due to different choices of $\nu_0$-values.

The stability at Z=114 in the calculations arises due to the Q-values
involved, as is evident from Fig. 3(a) where the calculated Q-values are
plotted as a function of the parent nuclear masses $^A$Z for both the
PCM and GLDM models, as well as that of another calculation by
Smola\'nczuk \cite{smolan97}. The Q-values are similar for the PCM and GLDM
models and are very small (minimum) for the $\alpha$-decay of $^{285}$114
parent. However, this is contrary to the predictions of Smola\'nczuk
\cite{smolan97}, which guided the Berkeley retracted experiment. Smola\'nczuk
\cite{smolan97} predicted the Q-values as an ever increasing function of the
parent nucleus charge (or mass), and became an apparent cause for the failure
of this experiment \cite{ninov99}.

Smaller Q-value should also mean a relative decrease in the penetrability P.
This is shown to be the case in our calculations presented in Fig. 3(b) where
$-log_{10}P$ vs. $^A$Z is plotted. It is further interesting to find that the
calculated preformation factor $P_0$ in Fig. 3(c) is also minimum for the
$\alpha$-nucleus preformation in $^{285}$114 parent. The $P_0$ factors in
present calculations are shown smaller by a few orders of magnitude
($\sim{10}^{-9}$), compared to that for the actinides \cite{balou99}.

So far, the Z=114 element (A=287-289) is synthesized only in hot fusion
reactions made at Dubna, using a $^{48}$Ca beam on $^{242,244}$Pu targets
\cite{oganess99b,oganess99c,oganess00}, which are found to result in larger
production cross sections \cite{hofmann00}. The cold synthesis of Z=114,
in the proposed reaction $^{76}Ge+^{208}Pb$ \cite{gupta77a}, is still
an open question experimentally where the measured cross section could throw
some light on its magic structure, if any. Apparently, the cold identification
of Z=114 will prove a testing ground for many structure calculations.

Figures 4 and 5 give the results of our calculations for heavy cluster decays
of each of the parents in the $\alpha$-decay chain of $^{293}$118, for a few
illustrative clusters. The choice of clusters is based on the minima in the
fragmentation potentials $V(\eta )$ and hence for the cases of largest
preformation factors $P_0$, illustrated as an example for Z=118 in Fig. 6.
The heaviest cluster included here in Figs. 4 and 5 is with Z=20
($^{48-50}$Ca) because the earlier calculations on PCM show that the two
processes of cold fission and cluster decay are almost indistinguishable for
clusters heavier than of mass $A_2>48$ \cite{kumar94}.

First of all we notice in Fig. 4(a) that the Q-value increases as
the size of cluster increases, but is almost independent of the
parent mass (the increase with parent mass is smooth, linear and
with a very small slope). However, the calculated cluster decay
half-lives, $log_{10}T_{1/2}^c$, in Fig. 4(b) present some
interesting results: (i) The decay half-lives for all the light
clusters, other than the $^{14}$C from Z=112-116 parents, are
rather high and hence the studied parents could be said as stable
against most of the light cluster decays. The shell stabilizing
effect, if any, is seen for $^{10}$Be decay of $^{289}$116 or
$^{273}$108 nucleus, since $T_{1/2}^c$ show strong peaking at
these two parent nuclei. (ii) The $^{14}$C decay, in particular
from $^{281}$112 parent, seems to present an interesting case of
(possibly, a reasonably) deformed magic daughter $^{267}$106 with
$N\approx 162$. This means to say that for the known deformed
magic shell at N=162 \cite{hofmann00}, the $^{267}$Sg, not yet
synthesized, could also be deformed and observed as $^{14}$C decay
of $^{281}$112 parent. (iii) The heavier clusters $^{48-50}$Ca are
predicted to decay with further smaller half-lives and hence
present themselves as further interesting cases of cluster decay
measurements. The calculated half-lives for $^{48-50}$Ca decays
lie far below the present limits of experiments, which go upto
$\sim 10^{28} s$ \cite{gupta94} for nuclei where enough atoms are
available. Here the closed shell effects of cluster (not daughter)
are playing the role, for which at present no measurements exist
in radioactive cluster decay studies. The heaviest cluster
observed so far is $^{32}$Si from $^{238}$Pu parent.

Figure 5 gives the cluster preformation ($P_0$) and penetration
(P) probabilities. Knowing that $T_{1/2}$ is a combined effect of
both $P_0$ and P ($\nu _0$ being almost constant), we notice in
Fig. 5 that though $^{10}$Be is better preformed (larger $P_0$)
than both $^{14}$C and Ca nuclei, but its penetration probabilty P
is so small that the $T_{1/2}$ for $^{10}$Be decay is much larger
than for either of the two other clusters. Thus, in experiments
one should consider the possibility of $^{14}$C and/ or $^{48}$Ca
decays in addition to $\alpha$-decay or fission of any of the
parents in $^{293}$118 $\alpha$-decay chain.

Summarizing, we have calculated the $\alpha$-decay of $^{293}$118 and its
subsequent parents ending the chain in $^{269}$106, as well as the heavy
cluster decays of all the parents in the $\alpha$-decay chain. Though the
experimental data for this $\alpha$-decay chain is retracted, the calculated
$\alpha$-decay half-lives are found to contain interesting nuclear structure
information. For the Q-values calculated from experimental binding energies,
supplemented by the Finite Range Droplet Model (FRDM) or the Thomas-Fermi
(TF) Model calculations, the $\alpha$-decay half lives show that there is a
magic shell structure at either or both Z=114, N$\approx$172, since the
calculated $T_{1/2}$ value for $\alpha$-decay of $^{285}$114 shows a strong
peaking structure. This is supported by a small Q-value and small (deeper
minima) penetrability and preformation factors for $\alpha$ in the
$^{285}$114 parent.

The cluster decay calculations for a decaying superheavy nucleus
formed in heavy ion reactions is made for the first time, with a
view to see if there is any branching of $\alpha$-decay to another
light nucleus due to the (spherical/ deformed) magicity of the
corresponding heavy daughter nucleus or due to the magicity of the
light nucleus itself. Interesting enough, the $^{293}$118 decay
chain offers two such possibilities: firstly, the $^{14}$C decay
of the inbetween parent $^{281}$112, and secondly, the doubly
magic $^{48}$Ca decay of any parent nucleus obtained after the
$\alpha$-decay(s) in the investigated chain. The first possibility
points out to the deformed magicity of the daughter product
$^{267}$106 at N=162 and the second possibility to a first time
observation of the doubly magic emitted cluster $^{48}$Ca. These
are exciting new possibilities for future studies.\\

\par\noindent
{\bf Acknowledgments:} \\
The authors are thankful to Professors Drs.
Gottfried M\"unzenberg and Sigurd Hofmann for many useful
discussions and for a careful reading of the manuscript. RKG, MB
and WS are thankful to Volkswagen-Stiftung, Germany, for the
support of this research work under a Collaborative Reserach
Project between the Panjab University and Giessen University. RKG
and MB are also thankful to the Council of Scientific and
Industrial Research (CSIR), New Delhi, for the partial support of
this research work. SK is thankful to Department of Atomic Energy
(DAE), Govt. of India for a Junior Research Fellowship.

\vfill\eject

\vfill\eject

\par\noindent
{\bf Figure Captions}\\

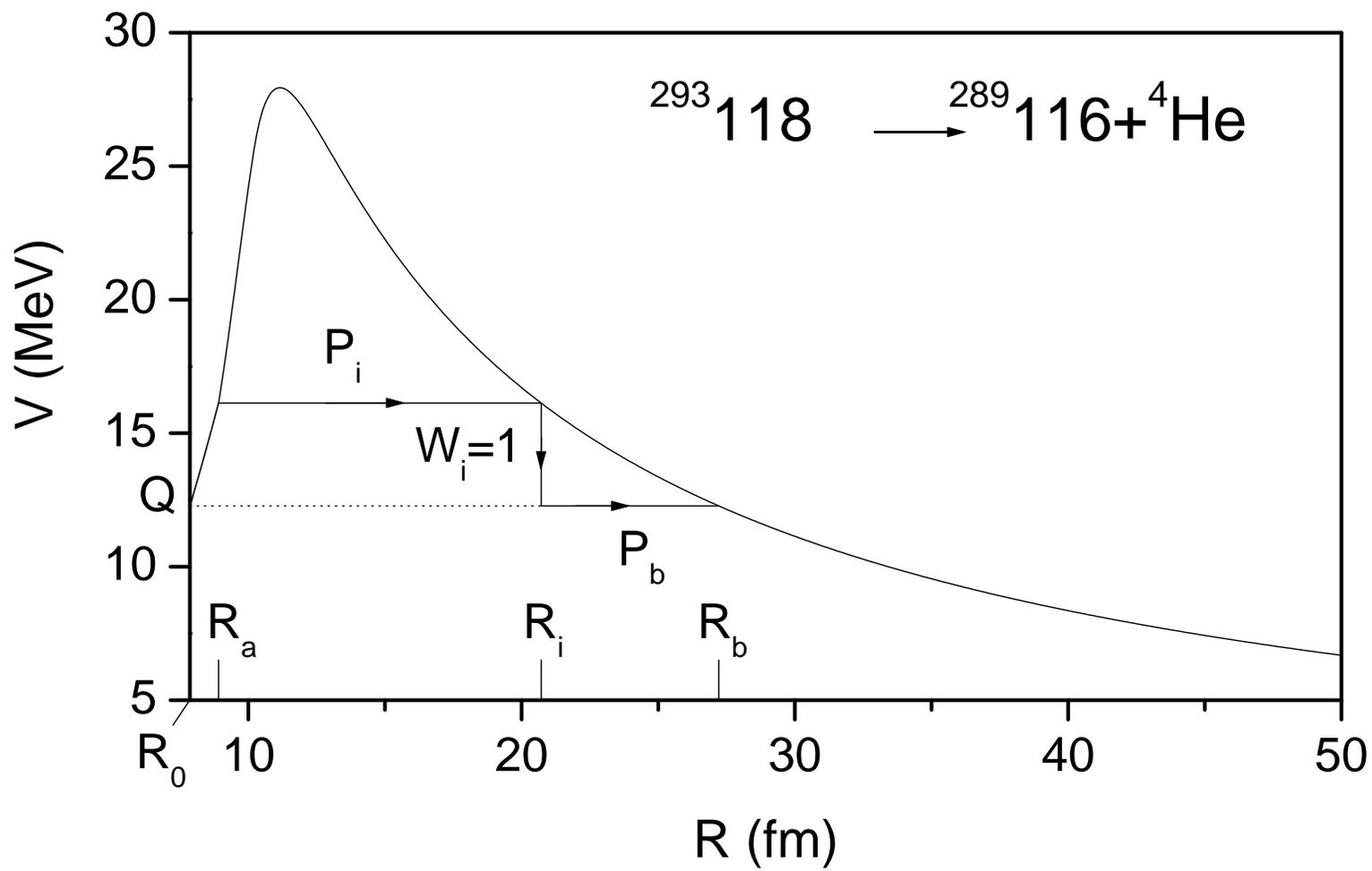
Fig. 1: The scattering potential for $\alpha$-decay of $^{293}$118, calculated
as the sum of Coulomb and nuclear proximity potential. The tunneling path used
by the PCM is also showm with the first and second turning point radii marked
as $R_a$ and $R_b$, respectively.\\

Fig. 2: The $\alpha$-decay half-lives calculated on PCM and compared with
GLDM model \cite{royer00} plotted as a function of the parent nucleus mass for
the $\alpha$-decay chain of $^{293}$118.\\ 

Fig. 3: (a) The Q-values calculated on FRDM \cite{moeller95}, compared with
TF \cite{myers96} and Smola\'nczuk \cite{smolan97} calculations, (b) the
penetration probabilty P, and (c) the preformation factor $P_0$ calculated on
PCM, plotted as a function the parent nucleus mass for the $\alpha$-decay
chain of $^{293}$118. \\

Fig. 4: The calculated Q-values on FRDM \cite{moeller95} and decay half-lives
on PCM for some cluster decays of the parents of $\alpha$-decay chain for
$^{293}$118 plotted as a function of the parent nucleus mass.\\

Fig. 5: Same as for Fig. 4, but for the penetration probabilty P and the
preformation factor $P_0$.\\

Fig. 6: (a) The fragmentation potential $V(\eta )$ for $^{293}$118, using the
experimental and FRDM binding energies \cite{audi95, moeller95}. Only half
of the potential as a function of the cluster mass $A_2$ is shown. The other
(symmetrical) half is a function of the daughter mass $A_1$.
Some of the minima are marked, showing the cluster mass.\\
(b) The preformation factor $P_0$, plotted as a function of both the cluster
and daughter masses, for the decaying $^{293}$118 parent.

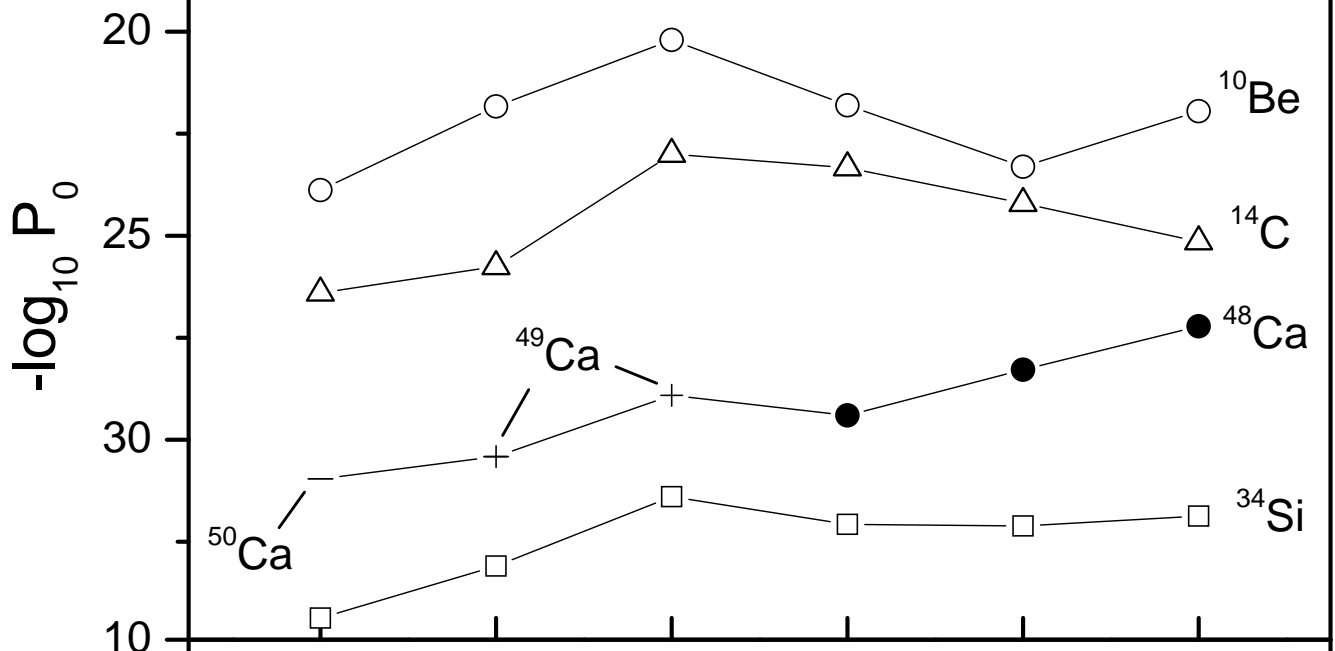
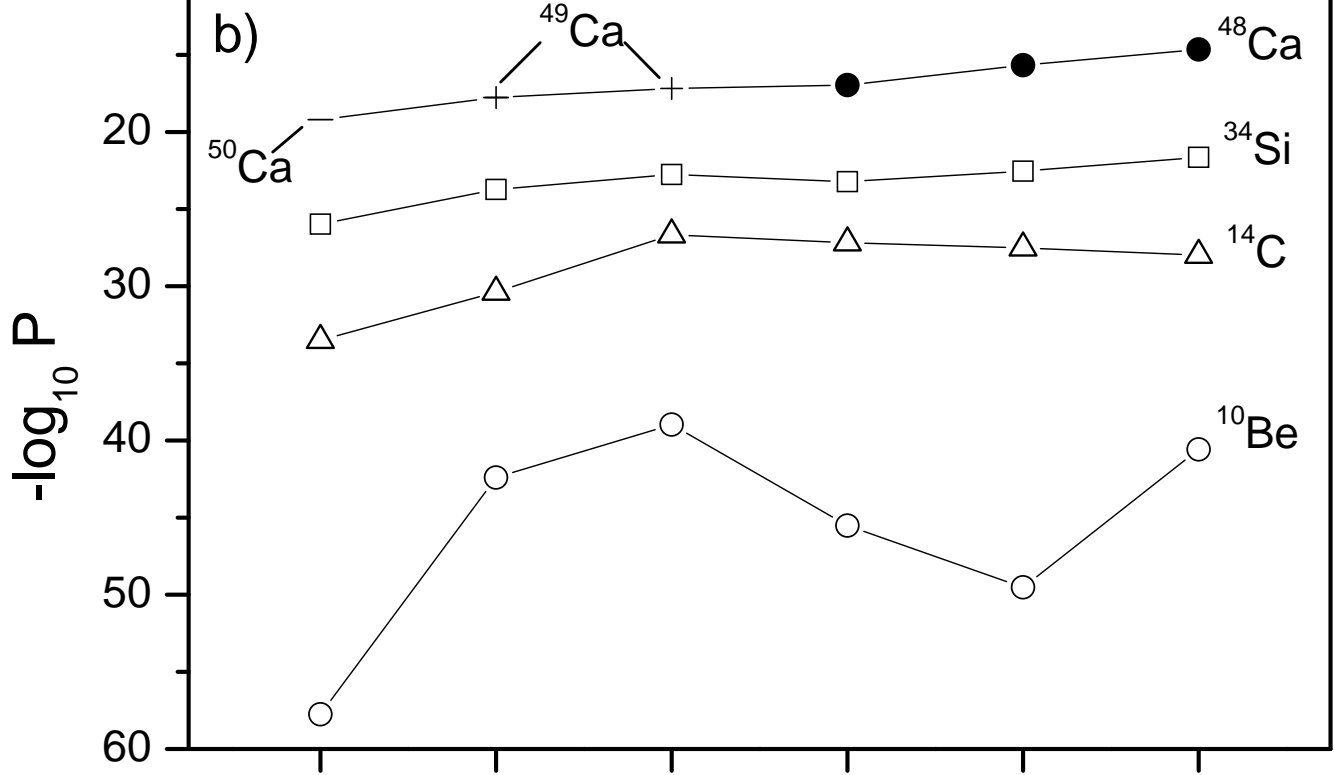
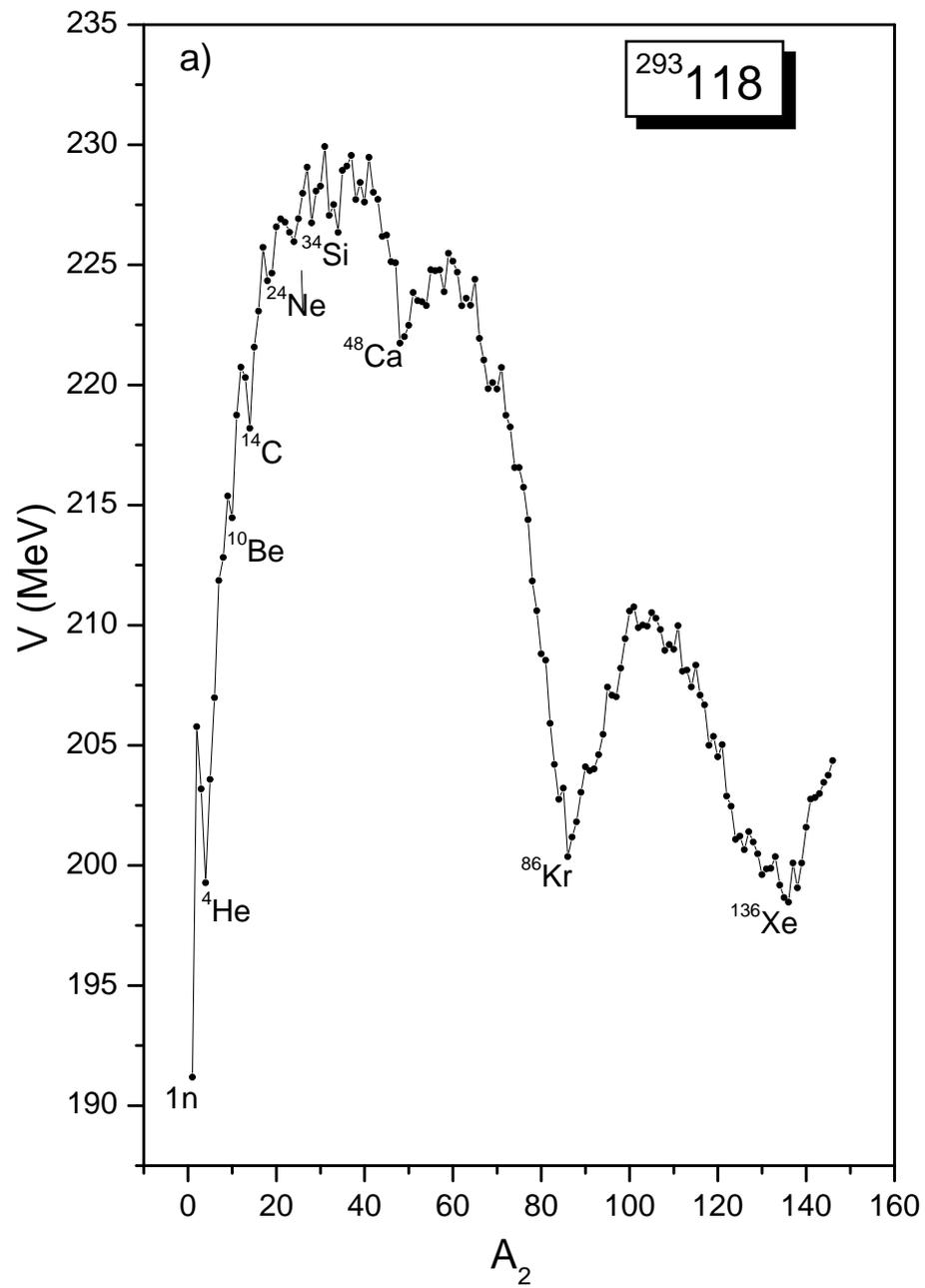
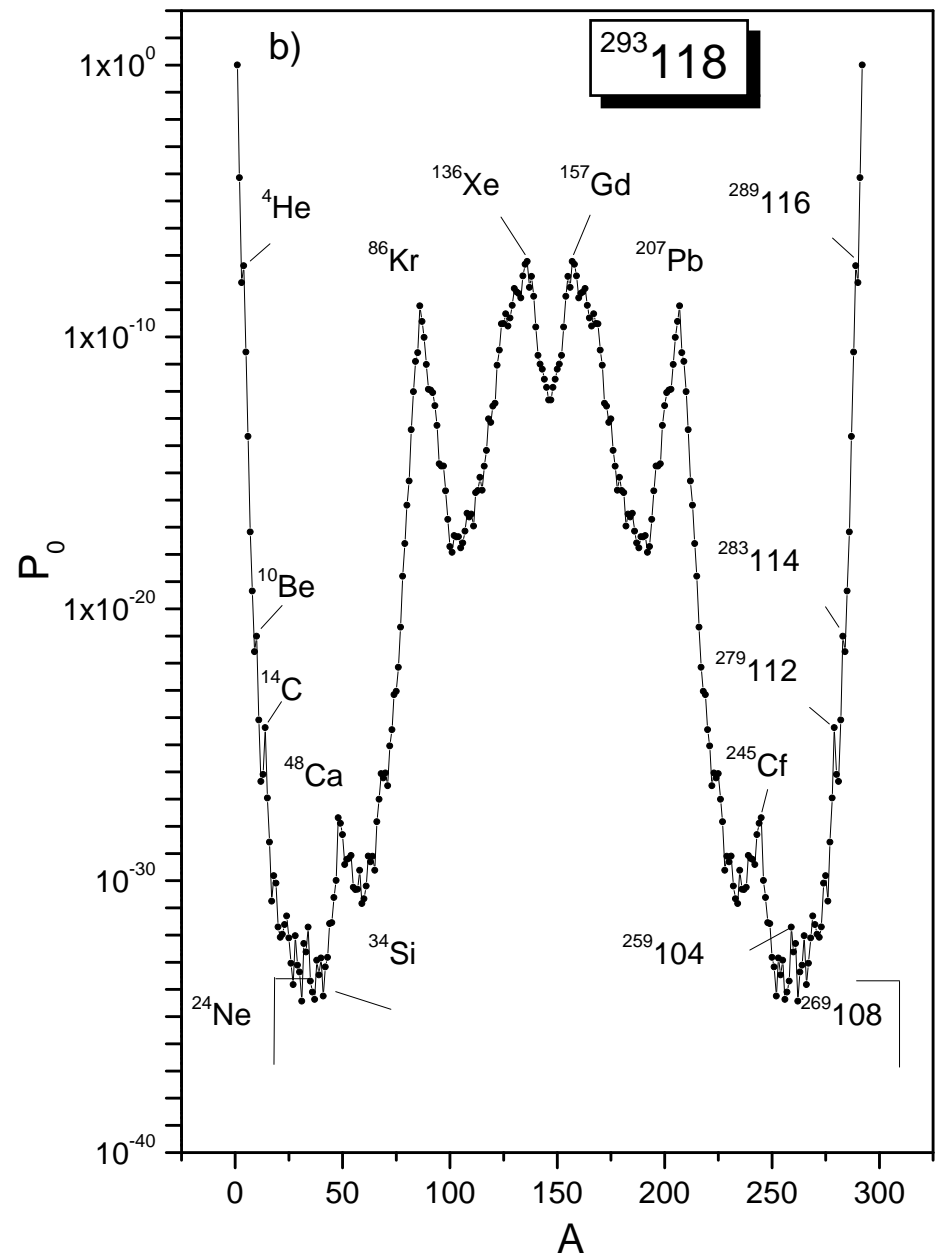
\end{document}